\documentclass[12pt]{article}

\usepackage{amssymb}
\usepackage{amsmath}
\usepackage[hiresbb]{graphicx}
\usepackage{amsthm}   %This is necessary for "proof"
\usepackage{authblk}   %This is necessary for author descriptions
\usepackage{lscape}   %This is necessary for "landscape"
\usepackage{url}
\usepackage{color}
\usepackage{here}

\usepackage{comment}
\usepackage{setspace}
\makeatletter

\@addtoreset{equation}{section}
\makeatother

\newtheorem{theo}{Theorem}
\newtheorem{assu}{Assumption}

\allowdisplaybreaks[4]

\newcommand{\indep}{\mathop{\perp\!\!\!\perp}}

\newcommand{\bld}{\boldsymbol}

\newcommand{\argmin}{\mathop{\rm arg~min}\limits}

\usepackage[top=1in,bottom=1in,left=1in,right=1in]{geometry}

\title{Robust Estimation and Model Selection for the Controlled Directed Effect with Unmeasured Mediator--Outcome Confounders}
\author[1]{Shunichiro Orihara \thanks{Address: 6-1-1 Shinjuku, Shinjuku-ku, Tokyo 160-8402, Japan; Email: orihara@tokyo-med.ac.jp}}
\author[2]{Shinpei Imori}
\author[3]{Kosuke Morikawa}
\author[4]{Atsushi Goto}
\author[1]{Masataka Taguri}

\affil[1]{Department of Health Data Science, Tokyo Medical University, Tokyo, Japan}
\affil[2]{Graduate School of Advanced Science and Engineering, Hiroshima University, Hiroshima, Japan}
\affil[3]{Graduate School of Engineering Science, Osaka University, Osaka, Japan}
\affil[4]{Department of Public Health, School of Medicine, Yokohama City University, Yokohama, Japan}

\date{}

\begin{document}
\begin{singlespace}
\maketitle
\end{singlespace}
\vspace{-1.5cm}
\section*{Abstract}
Controlled Direct Effect (CDE) is one of the causal estimands used to evaluate both exposure and mediation effects on an outcome. When there are unmeasured confounders existing between the mediator and the outcome, the ordinary identification assumption does not work. In this manuscript, we consider an identification condition to identify CDE in the presence of unmeasured confounders. The key assumptions are: 1) the random allocation of the exposure, and 2) the existence of instrumental variables directly related to the mediator. Under these conditions, we propose a novel doubly robust estimation method, which work well if either the propensity score model or the baseline outcome model is correctly specified. Additionally, we propose a Generalized Information Criterion (GIC)-based model selection criterion for CDE that ensures model selection consistency. Our proposed procedure and related methods are applied to both simulation and real datasets to confirm the performance of these methods. Our proposed method can select the correct model with high probability and accurately estimate CDE.

\vspace{0.5cm}
\noindent
{\bf Keywords}: double robustness, instrumental variable, model selection, structural mean model, UK biobank

\section{Introduction}
\begin{figure}[H]
\begin{center}
\includegraphics[width=16cm]{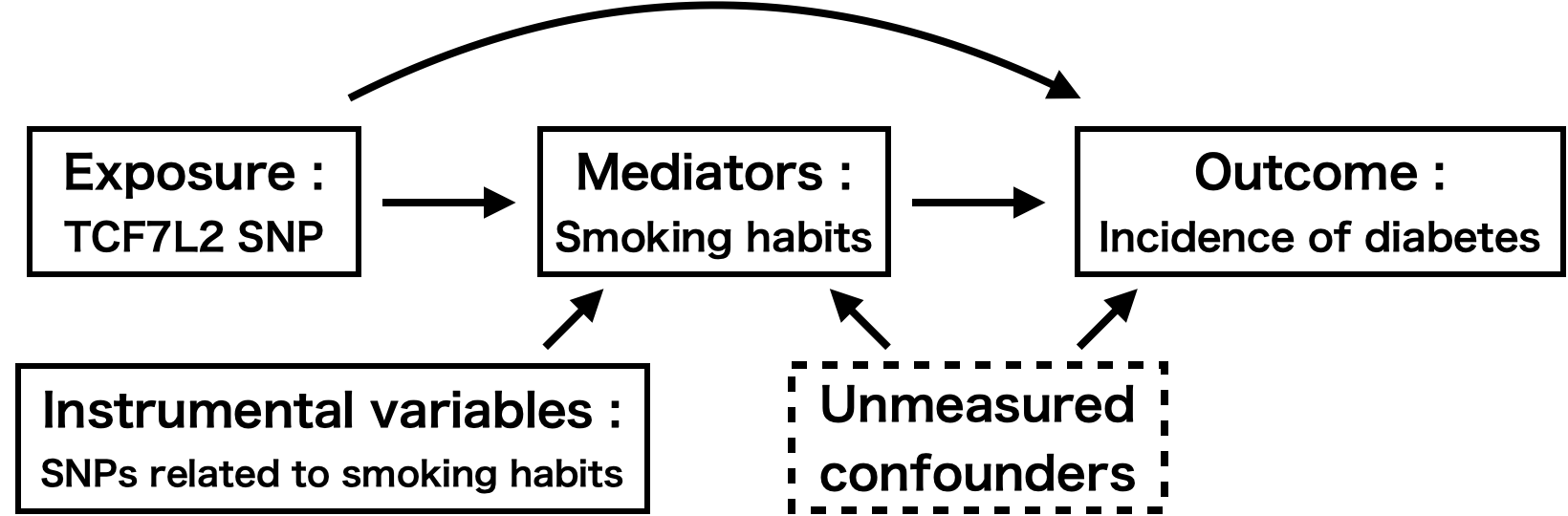}
\caption{Relationship of interested variables in this study with unmeasured confounders and instrumental variables}
\label{fig2}
\end{center}
\end{figure}

The estimation of causal effects and the exploration of their structures are central interests in causal inference. Causal mediation analysis meets these objectives by incorporating not only a treatment and an outcome variable but also the mediators that exist between them (Robins and Greenland, 1992; Pearl, 2022). In this manuscript, we explore the effect of a single nucleotide polymorphism (SNP) on the incidence of diabetes, considering smoking habits (measured by the number of cigarettes) as a mediator (refer to Figure \ref{fig2}).

Specifically, we focus on the TCF7L2 SNP (rs7903146; Chromosome 10), which has been previously reported to impair $\beta$-cell function (Cauchi et al., 2008), and may be directly associated with the incedence of diabetes. Since this SNP is considered the most significant genetic marker associated with diabetes incidence today (Vaquero et al., 2012), specifying the causal construction between the SNP and diabetes incidence is important. A recent study suggested a link between the SNP and nicotine poisoning in mice and rats (Duncan et al., 2019). However, its relevance in humans has been challenged by Lehrer and Rheinstein (2021). The volume of studies examining this relationship in humans is limited, necessitating further research for a definitive conclusion. Additionally, it is well-established that smoking habits are a risk factor for the incidence of diabetes (Pan et al., 2015).\\
\vspace{0.5cm}
\hspace{6.5cm}[Figure \ref{fig2} is here.]

To assess the causal effect, we focus on estimating the controlled direct effect (CDE; Pearl, 2022). CDE is defined as ``how much the outcome change on average if the mediator were fixed at level $m$ uniformly in the population but the treatment were changed from level $a^{*}=0$ to level $a=1$'' (page 23 of VanderWeele, 2015). CDE is thought to reflect the effect interpreted as the impact of fixing the mediator value to $m$ without intervening on the treatment (VanderWeele, 2009). Since we cannot directly manipulate the SNP but can modify smoking habits, this effect corresponds to our situation of interest. If CDE is small, it suggests that the SNP influences the incidence of diabetes primarily through its effects on smoking habits.

For accurate identification of CDE, confounders related to both the exposure--outcome and mediator--outcome must be observed and adjusted---an assumption known as `no unmeasured confounder.' Under this assumption, it is well-established that CDE can be identified (VanderWeele, 2015). Although SNPs are determined before birth and are often viewed as a form of random allocation (Burgess and Thompson, 2015), confounders still exist between smoking habits and the incidence of diabetes (Will et al., 2001; Campagna et al., 2019). In addition, thoroughly observing all confounders in observational studies remains a challenge; there is a concern that some unmeasured mediator--outcome confounders might exist. This issue is referred to as the `unmeasured confounder problem' in this manuscript.

Several methods have been well-considered to identify CDE together with the unmeasured confounder problem. Ten Have et al.\ (2007) introduced a straightforward outcome regression-based method anchored on the rank-preserving assumption. Joffe and Greene (2009) offered another outcome regression-based approach; however, it is distinguished from the Ten Have et al.\ method by its use of a surrogate outcome model (in our context, this is the mediator). Essentially, these methods require the assumption that exposure is independent of the (potential) outcome conditional on measured confounders (and mediator) to identify CDE. It is crucial to highlight that these methods demand accurate model specification and preclude the inclusion of an interaction term between the exposure and mediator in the outcome model. More recently, Sun and Ye (2020) and Wickramarachchi et al.\ (2023) have proposed a moment-based method. These approaches not only use the moments related to the outcome and mediator models but also the covariance between these models to identify the natural direct and indirect effects (Robins and Greenland, 1992; Pearl, 2022). Although these methods can identify CDE, they still require precise specification of both the outcome and mediator models.

In this manuscript, we propose an approach distinct from those in previous studies. We employ instrumental variables (IVs; Burgess and Thompson, 2015; Orihara et al., 2024) associated with the mediator to account for unmeasured mediator--outcome confounders. The concept of using IVs to adjust for unmeasured confounders has been considered in prior studies (e.g., Burgess et al., 2015; Rudolph et al., 2021; Chen et al., 2023). By leveraging these IVs, we construct an estimator based on an estimating equation for structured mean models (SMMs; refer to Vansteelandt and Joffe, 2014). Therefore, our method does not require the specification of all outcome models; only the parts of interest. Compared with previous methods, the risk of outcome model misspecification is decreased. Our proposed method requires valid IVs; however, it demonstrates double robustness: it remains a consistent CDE estimator if either the propensity score for the exposure or a subset of the outcome model is correctly specified---not necessarily all at once.

Also, to determine CDE precisely, we propose a Generalized Information Criterion (GIC)-type model selection criterion specifically tailored for the important components of the outcome model based on Nishii (1984). The ordinary GIC is known for its model selection consistency (Nishii, 1984), and we aim to assess its properties. Our proposed GIC-type model selection criterion retains consistency if a subset of the outcome model is correctly specified, regardless of whether the propensity score model is correctly specified.

Since many SNPs related to smoking habits have been identified in previous research (Erzurumluoglu et al., 2020; Lu et al., 2021), we use these SNPs as valid IVs to control for unmeasured mediator--outcome confounders in our study. Additionally, as the TCF7L2 SNP is presumed to be a random allocation, the propensity score model can be accurately specified, allowing us to derive a consistent CDE estimator and model selection criterion.

The remainder of this paper is organized as follows. From Sections 2 to 4, we propose novel estimation procedure and model selection criterion that exhibit certain robust properties. In Sections 5 and 6, we validate the properties of our proposed method through simulation experiments and analysis of real data obtained from the UK Biobank dataset. Supplementary material is provided in a separate supplemental file.

\section{Preliminaries}
Let $n$ denotes the sample size. Here, for $i = 1, \ldots, n$, $A_{i}\in\{0,1\}$, $\bld{M}_{i}\in\mathcal{M}$, $\bld{X}_{i}\in\mathcal{X}$, and $U_{i}\in\mathcal{U}$ represent a treatment, mediators, measured confounders, and unmeasured mediator--outcome confounders (referred to hereinafter as ``unmeasured confounder''), respectively. Additionally, we define potential outcomes $Y_{a\boldsymbol{m}i}\in\mathcal{Y}$ for each $a\in\{0,1\}$ and $\boldsymbol{m}\in\mathcal{M}$, as well as an observed outcome $Y_{i}$. It is assumed that the observed outcome $Y_{i}$ is consistent with the potential outcome $Y_{a\boldsymbol{m}i}$ when $A_{i}=a$ and $\boldsymbol{M}_{i}=\boldsymbol{m}$. Additionally, we introduce instrumental variables (IV) $\boldsymbol{Z}_{i}\in\mathcal{Z}$ associated with the mediators, with further details provided as we proceed. In our example, $A$ is TCF7L2 SNP, $M$ is smoking habits, $Y$ is an outcome related to the incidence of diabetes, and $Z$ is SNPs related to the smoking habits (see Figure \ref{fig2}), respectively.

Here, 
$$
\mathcal{M}\times\mathcal{X}\times\mathcal{Z}\times\mathcal{U}\times\mathcal{Y}\subset\mathbb{R}^{d_{M}}_{\geq0}\times\mathbb{R}^{d_{X}}\times\mathbb{R}^{d_{Z}}\times\mathbb{R}^{d_{U}}\times\mathbb{R}^{d_{Y}},
$$
where $\mathcal{M}$ is a finite set including $0$. Note that we assume that ${\rm Pr}(A=0,\bld{M}=\bld{0})>0$. Under these settings, we suppose that $(A_{i},\bld{M}_{i},\bld{X}_{i},\bld{Z}_{i},U_{i},Y_{i})$, $i=1,\, 2,\, \dots,\, n$ are {\it i.i.d.}\ copies from $(A,\bld{M},\bld{X},\bld{Z},U,Y)$. In this manuscript, we focus on estimating CDE defined as follows: for each $\bld{m}\in\mathcal{M}$,
$$
\Delta(\bld{m})={\rm E}\left[Y_{1\bld{m}}-Y_{0\bld{m}}\right]. 
$$

In the following discussion, we assume some independence assumptions. To avoid confusion, we summarize these assumptions here.
\begin{assu}{Independence Assumptions}\\
\label{assu1}
For each $a$ and $\bld{m}$,
\begin{gather}
Y_{a\bld{m}}\indep\bld{M}\mid A,\bld{X},U,\ \ Y_{a\bld{m}}\indep A\mid\bld{X},U, \label{ind1}\\
U\indep \bld{Z}\mid\bld{X},\ \ Y_{a\bld{m}}\indep\bld{Z}\mid A,\bld{X},U,\bld{M}, \label{ind2}\\
A\indep U\mid\bld{X},\ \ A\indep \bld{Z}\mid\bld{X},U. \label{ind3}
\end{gather}
\end{assu}\noindent
(\ref{ind1}) is necessary to identify CDE (VanderWeele, 2015), whereas (\ref{ind2}) relates to the ``valid'' IV conditions (Burgess and Thompson, 2015).

The IV necessaries to satisfy the following three conditions: 1) the IV is related to the mediators (corresponding to the regularity of matrix $\Gamma$ described in Theorem \ref{te1}), 2) the IV is not related to the unmeasured confounders (the former part of (\ref{ind2})), and 3) the IV is related to the outcome only through the mediators (the latter part of (\ref{ind2})).

(\ref{ind3}) is essential for advancing the discussion of our proposed method. The assumption is stated that an exposure variable needs to be independent of both unmeasured confounders and IVs related to mediators (conditional on some confounders). The former condition is identical to the standard requirement for ordinary IVs. In our context, where the variable is a SNP, this condition is presumed to be met. The latter condition relates to the independence assumption between SNPs. For example, if the exposure SNP and some SNPs related to the mediator belong to the same loci, this condition may not be satisfied (Burgess and Thompson, 2015). Therefore, the selection of SNPs associated with the mediator must be made with care. The similar independence assumption is considered in the context of valid IV selection (Sun et al., 2023).

\begin{assu}{No Additive Interaction (Wang and Tchetgen Tchetgen, 2018)}
\label{assu2}
\begin{align}
{\rm E}\left[Y_{a\bld{m}}-Y_{0\bld{0}}\mid\bld{X},U\right]={\rm E}\left[Y_{a\bld{m}}-Y_{0\bld{0}}\mid\bld{X}\right] \label{ind4}
\end{align}
\end{assu}\noindent
This assumption can be viewed as an extension of the ``rank preserving assumption'' (see, for example, Ten Have et al., 2007), and has been widely employed in various studies dealing with unmeasured confounders, such as the work by Wang et al.\ (2022) and Cui et al.\ (2023).

To estimate CDE $\Delta(\bld{m})$, we consider the conditional expectation
\begin{align}
\label{cond_mdl}
{\rm E}\left[Y-Y_{0\bld{0}}\mid A=a,\bld{X},\bld{M}=\bld{m},U\right].
\end{align}
To identify this quantity, we rely on (\ref{ind1}) in Assumption \ref{assu1} commonly assumed in ordinary mediation analysis. Under these independence assumptions, the quantity simplifies to
\begin{align*}
{\rm E}\left[Y-Y_{0\bld{0}}\mid A=a,\bld{X},\bld{M}=\bld{m},U\right]&={\rm E}\left[Y_{a\bld{m}}-Y_{0\bld{0}}\mid A=a,\bld{X},\bld{M}=\bld{m},U\right]\\
&={\rm E}\left[Y_{a\bld{m}}-Y_{0\bld{0}}\mid \bld{X},U\right].
\end{align*}
Furthermore, from (\ref{ind1}) in Assumption \ref{assu1} and (\ref{ind4}) in Assumption \ref{assu2}, the difference between the quantities, when averaged over $\bld{X}$, becomes CDE:
\begin{align}
\label{CDE1}
{\rm E}\left[{\rm E}\left[Y-Y_{0\bld{0}}\mid A=1,\bld{X},\bld{M}=\bld{m},U\right]-{\rm E}\left[Y-Y_{0\bld{0}}\mid A=0,\bld{X},\bld{M}=\bld{m},U\right]\right]\nonumber&\\
&\hspace{-12cm}={\rm E}\left[{\rm E}\left[Y_{1\bld{m}}-Y_{0\bld{0}}\mid \bld{X}\right]-{\rm E}\left[Y_{0\bld{m}}-Y_{0\bld{0}}\mid \bld{X}\right]\right]=\Delta(\bld{m}).
\end{align}

From the above discussion, it is natural to model the conditional expectation (\ref{cond_mdl}) as the following function of $a \in \{0, 1\}$, $\bld{m} \in \mathcal{M}$, and $\bld{x} \in \mathcal{X}$:
\begin{align}
\label{med_1}
 \bld{\tau}(a, \bld{m}, \bld{x})^\top \bld{\xi} = \bld{\tau}_1(a, \bld{m}, \bld{x})^\top \bld{\xi}_1 +  \bld{\tau}_2(\bld{m}, \bld{x})^\top \bld{\xi}_2, 
\end{align}
where $\bld{\xi} = {(\bld{\xi}_1^\top, \bld{\xi}_2^\top)}^\top$, $\bld{\tau}_1$ and $\bld{\tau}_2$ are known functions satisfying $\bld{\tau}_{1}(0,\bld{m},\bld{x})=\bld{0}$ and $\bld{\tau}_{2}(\bld{0},\bld{x})=\bld{0}$, ${(\bld{\xi}_1^\top, \bld{\xi}_2^\top)}^\top \in \bld{\Theta}_{\bld{\xi}}$, and $\bld{\Theta}_{\bld{\xi}}\subset\mathbb{R}^{L}$ is a parameter space to which $\bld{\xi}$ belongs. For instance, $\bld{\tau}_{1}(a,m,\bld{x})={a(1,m,m\bld{x}^{\top})}^{\top}$ and $\bld{\tau}_{2}(m,\bld{x})=m{(1,\bld{x}^{\top})}^{\top}$ with $d_{M}=1$. These two functions carry significant interpretations: $\bld{\tau}_{1}$ captures the interaction relationship between the treatment variable, mediators, measured confounders, and the outcome, while $\bld{\tau}_{2}$ represents the relationship between the mediator and the outcome under the reference treatment variable ($a=0$). We assume that there exists the true parameter $\bld{\xi}^0$ such that 
\begin{align*}
{\rm E}[Y - Y_{00} \mid A = a, \bld{X}, \bld{M} = \bld{m}, U] = \bld{\tau}(a, \bld{m}, \bld{X})^\top \bld{\xi}^0,
\end{align*}
where the superscript ``$0$'' on each parameter denotes the true values.

\section{Estimating Strategy and Double Robustness}
In this section, we address a scenario that necessitates the specification of parametric models for the propensity score and outcome. Our goal is to estimate CDE, which represents the target causal effect we are interested in. To achieve this, we use an estimating equation that includes IVs related to the mediator. We also explore the characteristics of the estimator with a focus on its ``double robustness.'' Although our primary data involves only a randomized treatment (i.e., the TCF7L2 SNP) and one mediator, we extend our consideration to more general situations, including those with multiple mediators and known confounders.

To estimate CDE under the model (\ref{med_1}), we would typically consider using an estimated equation based on the SMMs (Vansteelandt and Joffe, 2014). Unfortunately, we cannot apply this procedure directly due to the presence of some unmeasured confounders $U$. To address this issue, we use instrumental variables $\bld{Z}$ associated with the mediators, which satisfy (\ref{ind2}) in Assumption \ref{assu1}. Specifically, we propose a two-step estimation strategy: estimating 1) the parameters of propensity score, IVs, and the baseline outcome models (i.e., the models related to $Y_{0\bld{0}}$), and 2) the parameter $\bld{\xi}$ of model (\ref{med_1}).

\subsection{Working Models}
In the first step, we consider following two working models:
\begin{center}
$\mathcal{M}^{e}$: Propensity score model: $e(\bld{x};\bld{\alpha})$,\ \ \ $\mathcal{M}^{\varphi}$: Baseline outcome model: $\varphi(\bld{x};\bld{\zeta})$,
\end{center}
where $e: \mathcal{X}\to(0,1)$ and $\varphi: \mathcal{X}\to\mathbb{R}$. We assume that there exists the true parameters $\bld{\alpha}^{0}$ and $\bld{\zeta}^{0}$ such that ${\rm E}\left[A\mid \bld{X}_{i}\right]=e(\bld{X}_{i};\bld{\alpha}^{0})$ and
$$
{\rm E}\left[Y\mid A=0,\bld{X}_{i},\bld{M}=\bld{0},U\right]={\rm E}\left[Y_{0\bld{0}}\mid \bld{X}_{i},U\right]=\varphi(\bld{X}_{i};\bld{\zeta}^{0}).
$$
An estimator $\hat{\bld{\alpha}}$ for $\bld{\alpha}\in\bld{\Theta}_{\alpha}$ is obtained using procedures such as maximum likelihood estimation. Here, $\bld{\Theta}_{\alpha}$ is a parameter space to which $\bld{\alpha}$ belongs. Similarly, an estimator $\hat{\bld{\zeta}}$ for $\bld{\zeta}\in\bld{\Theta}_{\zeta}$ is obtained by solving the following estimating equation:
$$
\sum_{i=1}^{n}\dot{\varphi}(\bld{X}_{i};\bld{\zeta})\left(y_{i}-\varphi(\bld{X}_{i};\bld{\zeta})\right){\rm I}\left\{A_{i}=0,\bld{M}_{i}=\bld{0}\right\}=\bld{0},
$$
where $\dot{\varphi}(\bld{x};\bld{\zeta}):=\frac{\partial}{\partial\bld{\zeta}}\varphi(\bld{x};\bld{\zeta})$. Here, $\bld{\Theta}_{\zeta}$ is a parameter space to which $\bld{\zeta}$ belongs.

It is worth noting that model specification of $\varphi$ is commonly challenging, as the expectation includes unmeasured covariates $U$. One possibility is assuming that the expectation of the unmeasured covariates, conditioned on the observed values, becomes zero:
$$
{\rm E}\left[Y\mid A=0,\bld{X},\bld{M}=\bld{0},U\right]=\varphi(\bld{X};\bld{\zeta})+U,\ \ {\rm E}[U\mid \bld{X}]=0.
$$
Although it is hard to confirm the condition from the observed data, constructing the model is valuable. Our proposed procedure exhibits ``double robustness,'' meaning that the procedure yields a consistent estimator of CDE even if the baseline outcome model is misspecified.

\subsection{Double Robustness of Proposed Estimator}
In the second step, we focus on estimating the important parameter $\bld{\xi}$ in model (\ref{med_1}). Specifically, we consider the following estimating equation:
\begin{align}
\label{med_eq_1}
\sum_{i=1}^{n}\left(A_{i}-e(\bld{X}_{i};\hat{\bld{\alpha}})\right)\bld{Z}_{i}\left(y_{i}-\bld{\tau}_{i}^{\top}\bld{\xi}-\varphi(\bld{X}_{i};\hat{\bld{\zeta}})\right)=\bld{0},
\end{align}
where $\bld{\tau}_{i}\equiv \bld{\tau}(A_{i},\bld{M}_{i},\bld{X}_{i})$ (note that, henceforth, $d_{Z}=L$). Also, $\hat{\bld{\alpha}}$ and $\hat{\bld{\zeta}}$ are obtained from first step. To proceed with the discussions, we use (\ref{ind3}) in Assumption \ref{assu1} hereinafter.

The solution to (\ref{med_eq_1}), denoted as $\hat{\bld{\xi}}$, is given by
\begin{align}
\label{est1}
\hat{\bld{\xi}}&=\left(\sum_{i=1}^{n}\left(A_{i}-e(\bld{X}_{i};\hat{\bld{\alpha}})\right)\bld{Z}_{i}\bld{\tau}_{i}^{\top}\right)^{-1}\sum_{i=1}^{n}\left(A_{i}-e(\bld{X}_{i};\hat{\bld{\alpha}})\right)\bld{Z}_{i}\left(y_{i}-\varphi(\bld{X}_{i};\hat{\bld{\zeta}})\right).
\end{align}
Note that the inverse matrix exists asymptotically from a regularity condition {\bf C.4} in Appendix A of the supplemental file. Regarding the estimator, achieving consistency and asymptotic normality relies on the correctness of the two specified models. Hereafter, the phrase ``$\mathcal{M}^{(\cdot)}$ is correct'' indicates that the corresponding model is correctly specified. For instance, if ``$\mathcal{M}^{e}$ is correct,'' it means that the propensity score model is correctly specified; that is, $\bld{\alpha}^{*}=\bld{\alpha}^{0}$.
\begin{theo}{$\phantom{}$}\\
\label{te1}
Assuming that regularity conditions {\bf C.1}--{\bf C.6} in Appendix A of the supplemental file hold. If $\mathcal{M}^{e}$ or $\mathcal{M}^{\varphi}$ is correctly specified, then we have $\hat{\bld{\xi}}-\bld{\xi}^{0}=O_{p}(1/\sqrt{n})$. If $\mathcal{M}^{e}$ and $\mathcal{M}^{\varphi}$ are correctly specified, then we have
\begin{align}
\label{eq1}
\sqrt{n}\left(\hat{\bld{\xi}}-\bld{\xi}^{0}\right)\stackrel{L}{\to}N\left(\bld{0},\Gamma^{-1}\Sigma\left(\Gamma^{\top}\right)^{-1}\right),
\end{align}
where
$$
\Sigma={\rm E}\left[\left(A-e(\bld{X};\bld{\alpha}^{0})\right)^2\bld{Z}^{\otimes 2}\left(Y-\bld{\tau}^{\top}\bld{\xi}^{0}-\varphi(\bld{X};\bld{\zeta}^{0})\right)^2\right],\ \ \Gamma=-{\rm E}\left[\left(A-e(\bld{X};\bld{\alpha}^{0})\right)\bld{Z}\bld{\tau}^{\top}\right].
$$
\end{theo}\noindent
Here, $A^{\otimes 2}:=AA^{\top}$. The proof of Theorem \ref{te1} can be found in Appendix B of the supplemental file. The concept of double robustness here resembles the double robustness of the standard SMMs as discussed in Vansteelandt and Joffe (2014). However, the presence of unmeasured confounders and the use of IVs our approach apart from conventional SMMs; this robustness introduces a novel aspect to estimators within this context.

\section{Model Selection Criterion for CDE}
In this section, we introduce a GIC-based model selection criterion based on Nishii (1984) by modifying the penalty term of the information criterion proposed by Taguri et al. (2014) and Wallace et al. (2019). We also investigate its properties, particularly model selection consistency (Nishii, 1984; Shao, 1997).

\subsection{GIC-based Model Selection Criterion}
Here, we introduce additional notations. Following the notation used by Shao (1997), let $\mathcal{A}$ represents a set of candidate models, and $J\subset\{1, \ldots, p\}$ ($J\in\mathcal{A}$) represents a model. $\mathcal{A}^{c}$ denote the set of valid models, which is a subset of $\mathcal{A}$, and
$$
J^{0}=\left\{j=\{1,\dots,p\}\mid \xi_{j}^{0}\neq0\right\}
$$
denote the true model. Here, a ``valid model" refers to a model that includes the construction of the true model at the minimum. Thus, we have $J^{0}\in\mathcal{A}^{c}$, and $J^{0}$ has the minimum number of components among all $J\in\mathcal{A}^{c}$.

In Taguri et al.\ (2014) and Wallace et al.\ (2019), the following AIC-based information criterion is considered:
\begin{align*}
%\label{TAIC}
-2Q\left(\hat{\bld{\theta}}_{J}\right)+2 \times tr\left\{\widehat{\tilde{\Sigma}}_{J}\left(\widehat{\tilde{\Gamma}}_{J}^{-1}\right)^{\top}\right\},
\end{align*}
where $Q(\bld{\xi}_{J},\bld{\alpha},\bld{\zeta})\equiv Q(\bld{\theta}_{J})$ is a quasi-likelihood (QL):
\begin{align*}
Q(\bld{\theta}_{J})=-\frac{1}{2}\sum_{i=1}^{n}\left(A_{i}-e(\bld{X}_{i};\bld{\alpha})\right)\bld{\xi}_{J}^{\top}G_{J}^{\top}\bld{\tau}_{i}^{\otimes 2}G_{J}\bld{\xi}_{J}&\\
&\hspace{-5cm}+\sum_{i=1}^{n}\left(A_{i}-e(\bld{X}_{i};\bld{\alpha})\right)\bld{\xi}_{J}^{\top}G_{J}^{\top}\bld{\tau}_{i}\left(y_{i}-\varphi(\bld{X}_{i};\bld{\zeta})\right),
\end{align*}
$\bld{\xi}_{J}\in \mathbb{R}^{|J|}$ is a parameter under the model $J\in\mathcal{A}$ and $G_{J} \in \mathbb{R}^{p \times |J|}$ is a sub-matrix of ${\rm I}_p$ corresponding to $J$. For instance, when extracting the first $J$ component of $\bld{\tau}$, then $G_{J}=\left({\rm I}_{|J|},O\right)^{\top}$. Here, $\widehat{\tilde{\Sigma}}_{J}$ and $\widehat{\tilde{\Gamma}}_{J}$ are estimators for matrices $\tilde{\Sigma}_{J}$ and $\tilde{\Gamma}_{J}$ under the model $J\in\mathcal{A}$, where
\begin{align*}
\tilde{\Sigma}_{J}&={\rm E}\left[\left(\frac{\partial}{\partial\bld{\xi}_{J}}Q(\bld{\theta}_{J}^{*})\right)^{\otimes 2}\right]={\rm E}\left[\left(A-e(\bld{X};\bld{\alpha}^{*})\right)^2\left(G_{J}^{\top}\bld{\tau}\right)^{\otimes 2}\left(y-\bld{\tau}^{\top}G_{J}\bld{\xi}_{J}^{*}-\varphi(\bld{X}_{i};\bld{\zeta}^{*})\right)^2\right],\\
\tilde{\Gamma}_{J}&={\rm E}\left[\frac{\partial^{2}}{\partial\bld{\xi}_{J}^{\otimes 2}}Q(\bld{\theta}_{J}^{*})\right]=-{\rm E}\left[\left(A-e(\bld{X};\bld{\alpha}^{*})\right)\left(G_{J}^{\top}\bld{\tau}\right)^{\otimes 2}\right].
\end{align*}
Note that since the derivation of $-2Q(\bld{\xi}_{J},\hat{\bld{\alpha}},\hat{\bld{\zeta}})$ is not consistent with the left hand side of (\ref{med_eq_1}), the estimator (\ref{est1}) discussed in Section 3 may not minimize $-2 Q(\bld{\xi}_{J},\hat{\bld{\alpha}},\hat{\bld{\zeta}})$.

Here, to modify the penalty term of $QAIC_{CDE}$, we propose the GIC-based model selection criterion, denoted as ``$QGIC_{CDE}$'' (c.f., Nishii, 1984):
$$
QGIC_{CDE}(J):=-2Q\left(\hat{\bld{\theta}}_{J}\right)+\lambda_{n} \times tr\left\{\widehat{\tilde{\Sigma}}_{J}\left(\widehat{\tilde{\Gamma}}_{J}^{-1}\right)^{\top}\right\},
$$
where $\lambda_{n}$ is a deterministic sequence satisfying $\lambda_{n}\to\infty$ and $\lambda_{n}/n\to0\ (n\to\infty)$.

\subsection{Model Selection Consistency of Model Selection Criterion}
Under certain regularity conditions, the $QGIC_{CDE}$ exhibits a desirable property: model selection consistency
\begin{theo}{$\phantom{}$}\\
\label{te3}
Assuming that regularity conditions {\bf C.1}--{\bf C.8} in Appendix A of the supplemental file hold. If $\mathcal{M}^{\varphi}$ is correctly specified, then $QGIC_{CDE}$ exhibits model selection consistency; i.e.,
$$
{\rm Pr}\left(\hat{J}=J^{0}\right)\to 1,
$$
where $\hat{J}=\argmin_{J\in\mathcal{A}} QGIC_{CDE}(J)$.
\end{theo}\noindent
The proof of Theorem \ref{te3} is presented in Appendix B of the supplemental file. To best of our knowledge, Theorem \ref{te3} is the first derivation of the model selection criterion for CDE which has model selection consistency.

Finally, we conclude this section by discussing the model considerations related to the specific propensity score model used in our data analysis. As mentioned in Introduction, in our research the treatment variable (TCF7L2 SNP) is assumed to be randomly allocated. Therefore, the propensity score model can be correctly specified as $e(\bld{X}_{i};\hat{\bld{\alpha}})\equiv\frac{1}{n}\sum_{i=1}^{n}A_{i}=:\bar{A}$, which means implicity that the robustness property for parameter estimation holds exactly without considering the baseline outcome model. In the analysis presented in Section 5 and 6, we apply the statistics $\bar{A}$.

\section{Simulation Experiments}
In this section, we conduct simulation experiments to validate the proposed methods. Specifically, we aim to examine the performance of the proposed estimating procedure and model selection criteria, with particular emphasis on the double robustness. The iteration count for all simulation examples is set to 1000.

\subsection{Data-generating Mechanism}
First, we will explain the data-generating mechanism used in this simulation. We assume the presence of a single unmeasured confounder, denoted as $U_{i}\stackrel{i.i.d.}{\sim}N(0,2)$. Additionally, we assume that the assignment mechanism is completely random: $A_{i}\stackrel{i.i.d.}{\sim}Ber(0.5)$, and three instrumental variables (IVs) are denoted as $Z_{ki}\stackrel{i.i.d.}{\sim}Gamma(2,4)$, $k=1,2$. These variables are assumed to represent certain genetic variations.

Under this variable setting, a mediator has the following relationship:
\begin{align*}
\tilde{M}_{i}&\stackrel{i.i.d.}{\sim} Bin(4,p_{m}(A_{i},\bld{Z}_{i},X_{i},U_{i})),\\
p_{m}(A_{i},\bld{Z}_{i},X_{i},U_{i})&=expit\left\{-1+Z_{1i}+Z_{2i}+0.4U_{i}+0.4A_{i}\times U_{i}\right\},
\end{align*}
and $M_{i}=5\tilde{M}_{i}$. Therefore, the mediator takes on discrete values: ${0,5,10,15,20}$. The variable is assumed to represent the number of cigarettes smoked per day, and its value does not have a causal relationship with the exposure. Under this setting, the mean (SD) of the mediator is approximately $9.8$ $(8.7)$, $Cor(M,Z_{1}+Z_{2})\approx0.43$, and $Cor(M,U)\approx0.75$. The correlation between the mediator and the IVs is somewhat strong, but the objective of this simulation is to confirm the performance of methods under an ideal situation. In this sense, we consider the setting to be acceptable.

Finally, we specify the following outcome model: $Y_{00,i}=42+0.2U_{i}+\varepsilon_{yi}$, $\varepsilon_{yi}\stackrel{i.i.d.}{\sim}N(0,2)$, and $Y_{i}=2A_{i}+0.4M_{i}+Y_{00,i}$. The variable is assumed to represent the value of HbA1c, which serves as a surrogate for the incidence of diabetes (further details are provided in Section 6). Under this setting, there are only a few cases ($\approx$ 0\%) of diabetes (HbA1c $>$ 48) among the baseline subjects (i.e., $A=0$ and $M=0$), while there is a prevalence of 43\% of diabetes among all subjects.

\subsection{Model Candidates}
As the previous simulation experiment, we introduce three candidate outcome models to evaluate the performance of the model selection criteria. These models are as follows:
\begin{description}
\item[Model 1:] ${\rm E}\left[Y-Y_{00}\mid a,x,m,U\right]=a\xi_{1}+am\xi_{2}$,
\item[Model 2:] ${\rm E}\left[Y-Y_{00}\mid a,x,m,U\right]=a\xi_{1}+am\xi_{2}+m\xi_{3}$,
\item[Model 3:] ${\rm E}\left[Y-Y_{00}\mid a,x,m,U\right]=a\xi_{1}+am\xi_{2}+m\xi_{3}+mx\xi_{4}$.
\end{description}
Clearly, {\bf Model 2} represents the true model, while {\bf Model 3} includes the correct construction of the true model. Therefore, a model selection criterion that accurately selects either {\bf Model 2} or {\bf Model 3} is considered superior.

\subsection{Estimating Methods}
We considered three methods for estimating the causal effect. The proposed method, referred to as `Proposed,' was implemented as described in preceding sections. Two reference methods were employed in this study. The first, referred to as the `Ordinary SMM,' is based on the standard SMM with a model selection criterion proposed by Wallace et al.\ (2019) for a single time point (for further details, see Section 3.3 in Wallace et al., 2019). Since their criterion is of the AIC-type, we also explore a GIC-type model selection criterion by modifying the penalty term as previously described.

The second reference method, referred as the `2SLS,' utilizes the two-stage least squares (2SLS) estimator. For the mediator and outcome models, it adopts the approach outlined in Joffe and Greene (2009), without any model selection. Consequently, for 2SLS, we only consider cases where both the true outcome model and the true mediator model are correctly specified. Specifically, the 2SLS is a two-step procedure: firstly, regressing the instrumental variable (IV) and measured confounder on the mediator to derive the mediator's predictor; secondly, regressing the predicted mediator, exposure, and measured confounder on the outcome using the exact model specifications for Model 1. Since 2SLS meets the necessary assumptions and has shown satisfactory performance in the work of Joffe and Greene (2009), we use it as a benchmark in our study.

To estimate the parameters of interest ($\xi_{1}$ and $\xi_{2}$), it is essential to estimate the nuisance models, namely, the propensity score model (relevant only for the `Proposed' and `Ordinary SMM' methods) and the baseline outcome models. In this simulation, we only examine scenarios where the propensity score model is correctly specified. As a reference, we conduct model selection for the proposed method with unmeasured confounders (i.e., correctly specified baseline outcome model).

\subsection{Performance Metrics}
Regarding point estimates, the evaluation of the methods includes the calculation of the mean, empirical standard error (ESE), percent (\%) bias, root mean squared error (RMSE), and the visualization of estimated causal effects using boxplots in 1000 iterations. Bias and RMSE are calculated as $\% bias=100\left(\frac{\bar{\hat{\theta}}-\theta_{0}}{\theta_{0}}\right)$, and $RMSE=\sqrt{\frac{1}{1000}\sum_{k=1}^{1000}\left(\hat{\theta}_{k}-\theta_{0}\right)^2}$, respectively. Here $\bar{\hat{\theta}} = \frac{1}{1000}\sum_{k=1}^{1000}\hat{\theta}_{k}$ represents the average estimate over the 1000 iterations, $\hat{\theta}_{k}$ denotes the estimate obtained in each iteration, and $\theta_{0}$ is the true value of the parameter. $\theta_{0}$ corresponds to the true value of CDE under specific values of the mediator.

Regarding model selection, the number and proportion of selections that correctly identify the true model, or include the true model, out of the 1000 iterations are summarized.

\subsection{Simulation Results}
\begin{landscape}
\begin{figure}[H]
\begin{center}
\includegraphics[width=23cm]{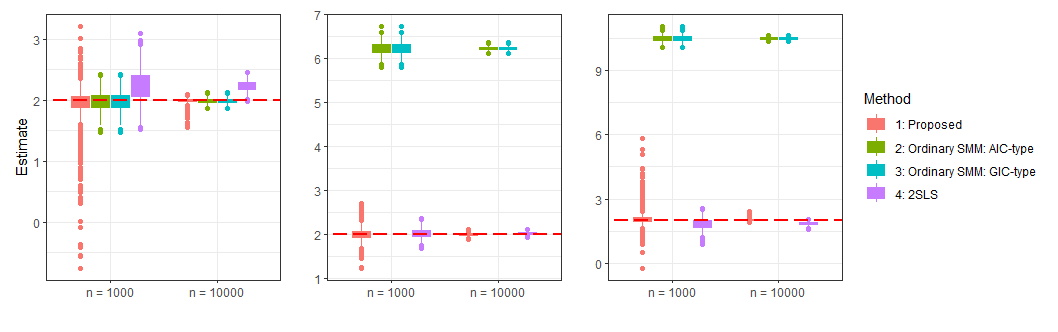}
\caption{Boxplots of estimates in 1000 iterations:\ The left panel: $CDE(0)$; the center panel: $CDE(10)$; The right panel: $CDE(20)$. Red dash line is the true value ($CDE(m)=2$)}
\label{fig3}
\end{center}
{\footnotesize{Ordinary SMM: Structural mean model with the model selection criterion proposed by Wallace et al.\ (2019),\\2SLS: 2SLS procedure regarding the mediator and the outcome model considered in Joffe and Grreene (2009),\\AIC-type: AIC type model selection criterion; the coefficient of the penalty term is set as ``$2$'',\\GIC-type: GIC type model selection criterion; the coefficient of the penalty term is set as ``$\log(n)$'',\\
$CDE(0)$: the value of the controlled direct effect (CDE) when the potential mediator value is $0$; $m=0$,\\
$CDE(10)$: the value of CDE when the potential mediator value is $10$; $m=10$\\
$CDE(20)$: the value of CDE when the potential mediator value is $20$; $m=20$\\
Regarding the results of the proposed method, there are some outliers. When $n=1000$, there are 8 results where the absolute value of $CDE(0)$ over 5; 1 result where the absolute value of $CDE(10)$ over 10; and there are 6 results where the absolute value of $CDE(20)$ over 15.
}}
\end{figure}
\end{landscape}

\begin{figure}[H]
\begin{center}
\includegraphics[width=18cm]{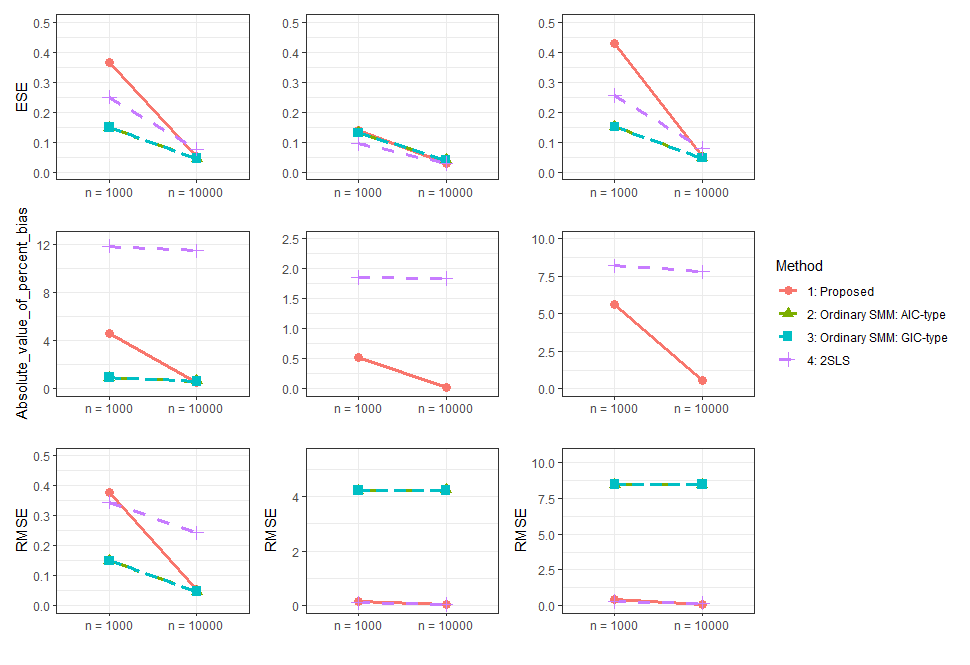}
\caption{Summary statistics of estimates:\ Empirical Standard Error (ESE), Absolute value of percent bias ($|\%$bias$|$), and Root Mean Squared Error (RMSE) in 1000 iterations were summarized. The left panel: $CDE(0)$; the center panel: $CDE(10)$; The right panel: $CDE(20)$.}
\label{fig4}
\end{center}
{\footnotesize{Ordinary SMM: Structural mean model with the model selection criterion proposed by Wallace et al.\ (2019),\\2SLS: 2SLS procedure regarding the mediator and the outcome model considered in Joffe and Grreene (2009),\\AIC-type: AIC type model selection criterion; the coefficient of the penalty term is set as ``$2$'',\\GIC-type: GIC type model selection criterion; the coefficient of the penalty term is set as ``$\log(n)$'',\\
$CDE(0)$: the value of the controlled direct effect (CDE) when the potential mediator value is $0$; $m=0$,\\
$CDE(10)$: the value of CDE when the potential mediator value is $10$; $m=10$\\
$CDE(20)$: the value of CDE when the potential mediator value is $20$; $m=20$\\
Since there are so large values, $|\%$bias$|$ of $CDE(10)$ and $CDE(20)$ for ordinary SMM are not shown.
}}
\end{figure}

The simulation results are summarized in Tables \ref{tab3}, \ref{tab4}, and Figure \ref{fig3} and \ref{fig4}. The results for Proposed method exhibit stability in terms of \%bias and RMSE for both $CDE(0)$, $CDE(10)$, and $CDE(20)$ when the sample size is large enough. In contrast, Ordinary SMM yields an unbiased baseline result ($CDE(0)$), but shows significant bias, especially for $CDE(20)$. These results can be attributed to the model selection. As evident from Table \ref{tab4}, the valid models cannot be selected, indicating obvious misspecification in model construction. Consequently, bias occurs when $m>0$. Regarding 2SLS, all estimates are slightly biased.

In our proposed method with correctly specified baseline outcome model, GIC-type has model selection consistency; this is consistent with the theoretical result. Whereas, in a misspecified situation, at least the underspecified model (Model 1) is not selected. In this sense, even if the baseline outcome model is misspecified, our proposed model selection criterion still works. For ordinary SMM, the model selection criteria do not work for all situations since estimates have obvious bias.

\vspace{0.5cm}
\hspace{4cm}[Table \ref{tab3}, \ref{tab4}, and Figure \ref{fig3} and \ref{fig4} are here.]

\section{UK Biobank Data Analysis}
We conduct a real data analysis using UK Biobank Data to examine the causal effect of the TCF7L2 SNP on the incidence of diabetes, mediated by smoking habits. It is widely recognized that diabetes affects specific organs such as the eyes, kidneys, and heart. Additionally, some cancers, such as colon cancer, have been associated with diabetes (Yu et al., 2022). With over 400 million individuals worldwide affected by diabetes (Cheng et al., 2019), prevention is crucial. To confirm the causal effect and underlying structure, we applied our proposed estimation strategy.

\subsection{Analysis Plan}
The mediator variable in our analysis is the number of cigarettes per day (NCPD; Data-Field: 2887), while the baseline HbA1c (HBBL; Data-Field: 30750) is used as the outcome variable. Additionally, Sex, baseline age (Data-Field: 21022), and baseline BMI (Data-Field: 21001) are considered as measured confounders between the mediator and outcome (Campagna et al., 2019). To handle the NCPD variable, any value over 20 is capped at 25, and responses such as ``Less than one a day'' or ``Do not know'' are treated as 0 to ensure accuracy in the analysis.

As evident from the previous sections, our model is applicable only to continuous outcomes. Therefore, in this study, we consider HBBL as a suitable surrogate for the incidence of diabetes in order to apply the model. Moreover, HBBL values are converted from IFCC units (the default unit in UK Biobank) to NGSP units for improved interpretability using the conversion formula: $0.09148 \times IFCC + 2.152$.

For our exposure variable (EX) and instrumental variables (IVs), we use SNPs available from the UK Biobank. The EX is coded as 0 in the absence of any variation from the wild type, and as 1 otherwise. For the IVs, we employ 77 SNPs previously identified as related to NCPD (Erzurumluoglu et al., 2020; Lu et al., 2021). Since EX is rs7903146 located on Chromosome 10, we do not use SNPs from Chromosome 10 as IVs to satisfy the second part of condition (\ref{ind3}).

In summary, our dataset comprises 502,419 participants with crude demographic or genetic data. For this study, we conducted a complete case analysis, meaning that only subjects with no missing data in the aforementioned variables were included in the analysis. Additionally, we excluded subjects with HBBL values exceeding 12, considering them as potential outliers (Landgra et al., 2022). Consequently, a total of 460,773 participants (91.7\% of the original sample) were included in the analysis.

We have prepared three candidate models: 1) including EX and the interaction term between EX and NCPD, 2) including EX and NCPD, or 3) including EX, NCPD, and the interaction term between EX and NCPD.

\subsection{Analysis Results}
Firstly, the demographic data of the participants are summarized in Table \ref{tab5}. It is evident that no significant trends are observed. Tables C.1--C.3 in Appendix C of the supplemental file summarize the correlations between the two variables. The correlation between SNP and NCPD was quite low, which may raise concerns about the weak IV problem.

Since EX is assumed to be allocated randomly, we can estimate the total effect (TE) simply: $0.0531\, (0.0490, 0.0572)$ (point estimate (99\% confidence interval (CI))). This indicates that subjects with a genetic mutation in the TCF7L2 SNP tend to have higher HbA1c levels, suggesting a negative causal effect of the TCF7L2 SNP on the incidence of diabetes. Using our proposed GIC-based model selection criterion under $\lambda_{n}=\log(n)$, Model 3 is selected. The estimated causal effects when NCPD is manipulated at levels of 0, 10, and 20 are summarized as follows (point estimate (99\% CI)): $-0.3874\, (-0.4150, -0.3694)$, $0.5231\, (0.4940, 0.5422)$, and $1.4337\, (1.3478, 1.4900)$, respectively. This indicates that by controlling the mediator, the causal effects change; intervening in smoking habits may reduce the risk of developing diabetes.

Additionally, we estimate the proportion eliminated (PE; VanderWeele, 2015). The definition is as follows: for all $m\in\mathcal{M}$,
$$
PE(m):=\frac{TE-CDE(m)}{TE}.
$$
In our analysis, $PE(0)$ and its 99\% CI is $8.2997\, (7.7800,8.8194)$. This indicates that nearly all of the TE is explained by the mediator. These results suggest that manipulating the number of cigarettes has an influence on the causal effect of the TCF7L2 SNP on diabetes incidence. Specifically, as the number of cigarettes consumed decreases, implying that quitting smoking, particularly intervening at $m=0$, can aid in diabetes prevention. 

Our data analysis has certain limitations. Firstly, the fact that the exposure, mediator, and outcome are measured at the same time point introduces the possibility of reverse causation (Burgess and Thompson, 2015). This temporal restriction poses a challenge in estimating causal effects accurately. Secondly, there are inherent risks associated with weak IVs. This concern is particularly relevant to the limitations of one-sample mendelian randomization (Burgess and Thompson, 2015). Although we attempt to address this issue by constructing allele scores, we cannot completely eliminate the concern. Furthermore, the confidence interval may be influenced by the presence of weak IVs. Consequently, it is crucial to interpret the results with awareness of the limitations of the analysis. Nevertheless, it is evident that our analysis has generated new insights: the effect of the SNP exposure on the incidence of diabetes can be modified by manipulating smoking habits.\\
\vspace{0.5cm}
\hspace{6.5cm}[Table \ref{tab5} is here.]

\section{Discussions and Conclusions}
In this manuscript, we propose a novel estimation and model selection procedure that exhibits double robustness for point estimation, and the model selection consistency. We mathematically establish these properties and validate them through simulation studies, comparing our approach with existing methods. Our findings suggest that our proposed methods outperform previous approaches, even when unmeasured confounders are absent. Additionally, in our analysis of UK Biobank data, we validate our research questions and conclude that the effect of SNP exposure on the incidence of diabetes can be modified by changing smoking habits.

While our proposed method offers several advantages, it does rely on certain assumptions---most notably, the strong assumption of parametric models for nuisance variables like the propensity score and the baseline outcome model. It is essential to relax this assumption, given that these models are not typically of primary interest and may not be universally applicable. One solution is to use nonparametric models. Fortunately, our method is based on estimating equations, as seen in (\ref{med_eq_1}), which share the same form as those in the context of double machine learning (Chernozhukov et al., 2018). By leveraging this approach, we can construct an estimation procedure that achieves $\sqrt{n}$--consistency for the target parameter of interest. We anticipate that model selection criteria based on this approach can also be derived.

Our proposed model selection criterion achieves model selection consistency if the baseline outcome model is correctly specified. However, as mentioned in Section 3, specifying the model may be challenging. When we observe some covariates, a considerably complex model is necessary to approximate the correct model. In other words, our proposed model selection criterion does not have ``double robustness." In Baba and Ninomiya (2021), they propose a doubly robust AIC-based model selection criterion, which implies that the model selection criterion becomes an asymptotically unbiased estimator of the risk when either the propensity score model or the outcome model is correctly specified.

It would be feasible to extend our estimation and model selection procedures to situations involving sequential treatments. In Wallace et al.\ (2019), these procedures were developed under the assumption of no sequential ignorability---an assumption that is often violated, especially in observational studies. More recent research, such as that by Cui et al.\ (2023), has considered situations involving sequential unmeasured confounders. By adapting our proposed methods, we aim to achieve not only accurate estimates of valid causal effects but also the identification of valid causal structures.

\vspace{1cm}
\noindent
{\bf Acknowledgements:} %We would like to express our grateful thanks to editors for their useful comments. This manuscript is sophisticated by their comments, and becomes more useful. We would like to thank Editage (www.editage.com) for English language editing. 
%\vspace{0.5cm}
This work was supported by JSPS KAKENHI Grant Number JP21K10500. All authors have accepted responsibility for the entire content of this manuscript and approved its submission.

%\vspace{0.5cm}
%\noindent
%{\bf Conflict of interest:} Authors state no conflict of interest.

\vspace{0.5cm}
\noindent
{\bf Data availability statement:} This study was conducted using the UK Biobank resource under application number 71079. The data that support the findings of this study are available from UK Biobank but restrictions apply to the availability of these data, which were used under license for the current study, and so are not publicly available. Data are however available from the authors upon reasonable request and with permission of UK Biobank. Simulation and Data analysis programs are available at the following URL:
\begin{itemize}
\item \url{https://github.com/SOrihara/DR_CDE}
\end{itemize}

\newpage

\begin{landscape}
\begin{table}[htbp]
\begin{center}
\caption{Summary of estimates: The iteration time is $1000$, and the true value of $CDE(0)$, $CDE(10)$, and $CDE(20)$ is $2$. Mean, empirical standard error (ESE), \%bias, and root mean squared error (RMSE) of estimated causal effects in 1000 iterations by estimating methods (``Estimating Method'' column) and the type of model selection criterions (``Model Selection Method'' column) are summarized.}
\label{tab3}
\begin{tabular}{ccc|cccc|cccc}\hline
{\bf Estimating}&{\bf Model Selection}&{\bf Parameters}&\multicolumn{4}{|c}{\bf Sample size: $n=1000$}&\multicolumn{4}{|c}{\bf Sample size: $n=10000$}\\
{\bf Method}&{\bf Method}&&{\bf Mean}&{\bf ESE}&{\bf $|\%$bias$|$}&{\bf RMSE}&{\bf Mean}&{\bf ESE}&{\bf $|\%$bias$|$}&{\bf RMSE}\\\hline
Proposed & GIC-type & CDE(0) & 1.908 & 0.366 & 4.577 & 0.377 & 1.990 & 0.052 & 0.497 & 0.053 \\ \cline{3-11}
 &  & CDE(10) & 2.010 & 0.140 & 0.517 & 0.140 & 2.000 & 0.029 & 0.019 & 0.029 \\ \cline{3-11}
 &  & CDE(20) & 2.112 & 0.429 & 5.610 & 0.443 & 2.011 & 0.051 & 0.534 & 0.052 \\ \hline
Ordinary SMM & AIC-type & CDE(0) & 1.981 & 0.148 & 0.954 & 0.150 & 1.987 & 0.045 & 0.634 & 0.047 \\ \cline{3-11}
 &  & CDE(10) & 6.226 & 0.131 & 211.296 & 4.228 & 6.231 & 0.040 & 211.572 & 4.232 \\ \cline{3-11}
 &  & CDE(20) & 10.471 & 0.151 & 423.546 & 8.472 & 10.476 & 0.046 & 423.777 & 8.476 \\ \cline{2-11}
 & GIC-type & CDE(0) & 1.981 & 0.148 & 0.954 & 0.150 & 1.987 & 0.045 & 0.634 & 0.047 \\ \cline{3-11}
 &  & CDE(10) & 6.226 & 0.131 & 211.296 & 4.228 & 6.231 & 0.040 & 211.572 & 4.232 \\ \cline{3-11}
 &  & CDE(20) & 10.471 & 0.151 & 423.546 & 8.472 & 10.476 & 0.046 & 423.777 & 8.476 \\ \hline
2SLS & - & CDE(0) & 2.238 & 0.249 & 11.908 & 0.345 & 2.230 & 0.077 & 11.495 & 0.242 \\ \cline{3-11}
 &  & CDE(10) & 2.037 & 0.094 & 1.857 & 0.101 & 2.037 & 0.029 & 1.841 & 0.047 \\ \cline{3-11}
 &  & CDE(20) & 1.836 & 0.255 & 8.194 & 0.303 & 1.844 & 0.078 & 7.812 & 0.175 \\ \hline
\end{tabular}
\end{center}
{\footnotesize{Ordinary SMM: Structural mean model with the model selection criterion proposed by Wallace et al.\ (2019),\\2SLS: 2SLS procedure regarding the mediator and the outcome model considered in Joffe and Grreene (2009),\\AIC-type: AIC type model selection criterion; the coefficient of the penalty term is set as ``$2$'',\\GIC-type: GIC type model selection criterion; the coefficient of the penalty term is set as ``$\log(n)$'',\\
$CDE(0)$: the value of the controlled direct effect (CDE) when the potential mediator value is $0$; $m=0$,\\
$CDE(10)$: the value of CDE when the potential mediator value is $10$; $m=10$\\
$CDE(20)$: the value of CDE when the potential mediator value is $20$; $m=20$
}}
\end{table}
\begin{table}[htbp]
\begin{center}
\caption{Summary of the number and proportion of selecting valid models: The iteration time is $1000$. The number and proportion of selections that correctly identify the true model, or include the true model in 1000 iterations by estimating methods (``Estimating Method'' column) and the type of model selection criterions (``Model Selection Method'' column) are summarized.}
\label{tab4}
\begin{tabular}{cc|cc|cc}\hline
{\bf Estimating}&{\bf Model Selection}&\multicolumn{2}{|c}{\bf Sample size: $n=1000$}&\multicolumn{2}{|c}{\bf Sample size: $n=10000$}\\
{\bf Method}&{\bf Method}&&{\bf Including}&&{\bf Including}\\
&&{\bf Correct Model}&{\bf Correct Model}&{\bf Correct Model}&{\bf Correct Model}\\\hline
Proposed&GIC-type&747 (74.7)&919 (91.9)&914 (91.4)&1000 (100)\\\hline
Ordinary&AIC-type&0 (0.0)&0 (0.0)&0 (0.0)&0 (0.0)\\\cline{2-6}
SMM&GIC-type&0 (0.0)&0 (0.0)&0 (0.0)&0 (0.0)\\\hline\hline
\begin{tabular}{c}Reference: Proposed with\\Unmeasured Confounder\end{tabular}&GIC-type&760 (76.0)&925 (92.5)&967 (96.7)&1000 (100)\\
\hline
\end{tabular}
\end{center}
{\footnotesize{Ordinary SMM: Structural mean model with the model selection criterion proposed by Wallace et al.\ (2019),\\AIC-type: AIC type model selection criterion; the coefficient of the penalty term is set as ``$2$'',\\GIC-type: GIC-type model selection criterion; the coefficient of the penalty term is set as ``$\log(n)$''
}}
\end{table}
\end{landscape}

\begin{table}[htbp]
\caption{Summary of demographic characteristics}
\label{tab5}
\begin{center}
\begin{tabular}{c|ccc}\hline
&Exposure group&Non-exposure group&Total\\
&$N_{1}=227965$&$N_{0}=232808$&$N=460773$\\\hline
Sex&104620 (45.9\%)&106121 (45.6\%)&210741 (45.7\%)\\\hline
Age&56.6 (8.10)&56.5 (8.09)&56.5 (8.09)\\\hline
BMI&27.4 (4.74)&27.5 (4.82)&27.4 (4.78)\\\hline
NCPD&3.96 (7.80)&3.97 (7.82)&3.97 (7.81)\\\hline
\end{tabular}
\end{center}
{\footnotesize{Continuous variables are summarized as: $Mean\, (SD)$.\\
Category variables are summarized as: $Frequence\, (Percent)$.\\
NCPD: The number of cigarettes per day
}}
\end{table}

\newpage

\newpage

\newpage

\end{document}